\documentclass[letterpaper,journal]{IEEEtran}
\usepackage{graphicx} % Required for inserting images
\usepackage{cite}
\usepackage{soul,color}
\usepackage{multirow}
\usepackage{url}
\usepackage{booktabs}
\usepackage{tikz}
\usepackage{subcaption}
\usepackage{xcolor}
\usepackage{float}
\usepackage[normalem]{ulem}
\usepackage{amsmath, MnSymbol}

\newcommand{\ballnumberwhite}[1]{\tikz[baseline=(myanchor.base)] \node[circle,fill=white,draw=black,inner sep=1pt] (myanchor) {\color{black}\bfseries\footnotesize #1};}

\title{Anomaly-Flow: A Multi-domain Federated Generative Adversarial Network for Distributed Denial-of-Service Detection}

\IEEEpubid{\begin{minipage}{\textwidth}\ \\[12pt]
\\
\\
Leonardo Henrique de Melo, Gustavo de Carvalho Bertoli, and Louren\c{c}o \\ Alves Pereira Junior are with the Aeronautics Institute of Technology (ITA). \\E-mails: \{leonardomelo, bertoli, ljr\}@ita.br. Michele Nogueira and Aldri \\Santos are with the Federal University of Minas Gerais, Department of \\Computer Science. E-mails: \{michele,aldri\}@dcc.ufmg.br
\end{minipage}}

\author{Leonardo Henrique de Melo, Gustavo de Carvalho Bertoli, Michele Nogueira,\\ Aldri Luiz dos Santos, Louren\c{c}o Alves Pereira Junior}
\date{January 2024}

\begin{document}

\maketitle

%TC:ignore
\begin{abstract}
Distributed denial-of-service (DDoS) attacks remain a critical threat to Internet services, causing costly disruptions. While machine learning (ML) has shown promise in DDoS detection, current solutions struggle with multi-domain environments where attacks must be detected across heterogeneous networks and organizational boundaries. This limitation severely impacts the practical deployment of ML-based defenses in real-world settings.

This paper introduces Anomaly-Flow, a novel framework that addresses this critical gap by combining Federated Learning (FL) with Generative Adversarial Networks (GANs) for privacy-preserving, multi-domain DDoS detection. Our proposal enables collaborative learning across diverse network domains while preserving data privacy through synthetic flow generation. Through extensive evaluation across three distinct network datasets, Anomaly-Flow achieves an average F1-score of $0.747$, outperforming baseline models. Importantly, our framework enables organizations to share attack detection capabilities without exposing sensitive network data, making it particularly valuable for critical infrastructure and privacy-sensitive sectors.

Beyond immediate technical contributions, this work provides insights into the challenges and opportunities in multi-domain DDoS detection, establishing a foundation for future research in collaborative network defense systems. Our findings have important implications for academic research and industry practitioners working to deploy practical ML-based security solutions.
\end{abstract}
%TC:endignore

\begin{IEEEkeywords}
Multi-domain, DDoS, Federated Learning, GAN, Network Attacks, Anomaly Detection
\end{IEEEkeywords}

\section{Introduction}
\label{sec:introduction}

Learning across multiple domains is challenging for machine learning (ML) applications~\cite{generalization-survey}, especially in network environments with diverse segments and heterogeneous participants. 
An adversary can exploit this segmented characteristic to launch attacks without awareness by all segments once a global network view is challenging from a defense perspective. Federated Learning (FL) is an ML technique that helps to overcome this segmented characteristic through collaborative attack detection while maintaining privacy~\cite{AGRAWAL2022346}.

From a myriad of network attacks, Distributed Denial of Service (DDoS) is an impactful attack that aims to deny services by flooding computational resources with many requests. 

Fortunately, recent research has introduced FL-based mechanisms to detect DDoS attacks, enabling learning across different networks and distributed DDoS targets~\cite{flddos, fleam, FLAD2024}. However, it is not common practice for researchers to evaluate their proposals in a multi-domain perspective~\cite{xenids}. Currently, using ML techniques for DDoS detection is ineffective in an operational environment~\cite{CATILLO2021102341}, limiting deployment. Among other reasons, ML algorithms cannot perform satisfactorily in different network domains~\cite{Verkerken2021}, challenging this transposition from research to operation. Thus, evaluating the model’s effectiveness across multiple domains is essential to ensure that it accurately captures the underlying behavior of DDoS attacks. An approach for ML models to perform satisfactorily would be sharing data from these multi-domains~\cite{generalization-survey}. However, this approach leads to other problems, such as loss of privacy and lack of data.

In this work, to improve the multi-domain DDoS detection, we propose using FL as an applicable technique for detecting network attacks across multi-domains~\cite{cose}.
As data sharing between domains is helpful but challenging, we integrate Generative Adversarial Networks (GANs) to generate synthetic network data, allowing its sharing between domains without compromising privacy~\cite{netshare}. 
Then, this study proposes a multi-domain DDoS detection method that leverages FL and GAN-based synthetic data to address privacy constraints and domain generalization, resulting in the proposed Anomaly-Flow.

Our evaluation found that Anomaly-Flow improves DDoS detection across multiple networks using different datasets. Using the integrated GANs’ ability to generate synthetic data, Anomaly-Flow enables learning DDoS patterns between various domains. Additionally, heterogeneous models can be shared with external entities with lower privacy concerns, and the learned DDoS behaviors can be shared with those not part of the FL scheme. To our knowledge, Anomaly-Flow is the first method for multi-domain DDoS detection, and it considers three steps: FL DDoS detection, synthetic data generation, and heterogeneous model sharing. 

In the upcoming sections, we explore ML for DDoS detection and multi-domain generalization. We detail a use case involving ML techniques, including the design and implementation of Anomaly-Flow. We also analyze the performance of our approach in diverse network settings and discuss the implications for network security advancement. Further, DDoS detection between multiple domains is an open challenge. Therefore, we outline the challenges and opportunities identified during our experiments. We conclude with reflections on our study and future research directions for advancing multi-domain DDoS detection.

\section{DDoS Detection and Multi-domain Evaluation}\label{sec:related}

This section presents previous work applying ML, specifically GANs and FL, to DDoS detection. Our analysis evaluates these works from the multi-domain detection perspective, typically assessed through a cross-evaluation between different datasets. Overall, multi-domain DDoS detection has received limited attention in the literature. Then, we discuss related works that present multi-domain detection when applying ML to flow-based network data for security-related tasks.

Using GANs for network flows, \cite{netshare} proposed the Netshare framework for the synthetic data generation of IP headers and specific network flow attributes. The authors’ proposed solution has three aspects: ``fidelity'', ``scalability-fidelity tradeoff'' and ``privacy-fidelity tradeoffs.''
The authors stated that the proposed framework, considering proximity metrics between synthetic and actual data, obtained 46 percent more fidelity than the baseline.
Furthermore, the authors showed that GAN presents the best tradeoff between scalability and fidelity among all their baselines. 
Despite their seminal contribution of applying GANs to network flow data, the discussion about anomaly detection and the experiments conducted are superficial, which can hinder the reproducibility of the results. Additionally, no specific considerations for DDoS are presented.

Regarding FL applied to DDoS detection, \cite{flddos} proposed FLDDoS as a federated recurrent neural network (RNN) model to detect malicious network activities and maintain privacy. The authors proposed hierarchical clustering to aggregate local models using the k-means algorithm and a method to balance the number of benign and attack samples among the clients. Furthermore, the authors used an Autoencoder (AE) to extract features automatically during the training rounds. However, the authors did not evaluate the detection across multi-domains and the model’s capability to identify DDoS on unseen data.  
Thus, the evaluation of FLDDoS in a multi-domain scenario is not presented.

Also applying FL for DDoS, \cite{PST2022} presented FL for DDoS detection in a multi-tenant scenario. This scenario represents the challenge of DDoS attacks against a specific tenant going unaware by other tenants. To overcome this challenge, they proposed the use of FL. The multi-tenant setting is similar to the multi-domain, but their proposed methodology focused only on a tenant’s ability to learn about DDoS attacks from other tenants. The authors used the same dataset, divided between different tenants for FL training and evaluation. It fails to represent a comprehensive evaluation across multiple domains, which would require different datasets to represent other network domains.

The work \cite{FLAD2024} also focused on FL applied to DDoS detection. It presented a form of aggregation based on the performance of federated clients. The authors validated the proposed architecture using an MLP model to detect multiple classes of DDoS attacks in a dataset. Compared with FLDDoS~\cite{flddos} and traditional FedAvg, the results demonstrated better convergence time and performance considering ML learning metrics. Furthermore, the authors evaluated the traffic classification of unseen data using the model. However, similar to \cite{PST2022}, the presented approach evaluated different attacks from the exact origin without considering a multi-domain evaluation. Additionally, the authors did not explore strategies for multi-domain performance through information sharing or training heterogeneous models with external entities.

Addressing evaluations in a multi-domain setting, \cite{xenids} presented the XeNIDS framework for cross-evaluation between multiple datasets. The authors highlighted the challenges of deploying anomaly-based NIDS in real-world networks, noting that while many studies achieve near-perfect results, these outcomes often do not translate well into practical applications due to this lack of multi-domain evaluation. The authors proposed a methodology for training and testing models with various attack types. They identified ten contexts for cross-evaluation based on the presence or absence of attack types and benign data during the training and testing phases. This multi-context approach can aid in generating augmented data. However, there are significant concerns regarding data privacy, as different data compositions can lead to data leaks or make practical implementation challenging due to privacy issues between multiple parties.

Then, \cite{efc} proposed the Energy-based Flow Classifier (EFC) algorithm for identifying malicious network flow data. The authors tested the model’s generalization capability using three datasets: CIDDS-001, CICIDS-2017, and CIC-DDoS2019. The binary classification task involves determining whether flows are benign or anomalous.
The algorithm, used to identify anomalies in graph structures, demonstrated promising results in a cross-evaluation analysis between two datasets involving the same attack type, representing a multi-domain setting. However, the evaluation is limited to similar datasets and does not focus on DDoS. The authors suggested extending it to assess performance across distinct datasets.

Lastly, \cite{Verkerken2021} employed different ML algorithms for anomaly detection, including Principal Component Analysis (PCA), Isolation Forest (IF), AE, and One-Class Support Vector Machine (oSVM). The authors conducted cross-evaluation between two different datasets containing similar types of attacks. The results showed that the anomaly detection models performed poorly in this multi-domain setting, highlighting the need for adaptable and robust models. Additionally, it is a limitation evaluating closely related datasets as \cite{efc}, no focus on DDoS was given, and the work did not discuss the performance degradation in diverse network topologies with more heterogeneous flows.

In summary, we identified gaps in DDoS detection research involving FL and GANs, including limited use of GANs for DDoS-specific synthetic data \cite{netshare}, FL approaches \cite{flddos, PST2022, FLAD2024} lacking multi-domain generalization and sharing beyond participants, and insufficient privacy consideration in multi-domain settings \cite{xenids} and \cite{efc}.
Anomaly-Flow addresses these gaps by (1) integrating FL and GANs for multi-domain DDoS detection across heterogeneous networks while preserving privacy, (2) generating shareable synthetic network flows, and (3) enabling external entities to benefit from collaborative learning via heterogeneous model sharing without compromising data. This is the first framework to combine these features for practical multi-domain DDoS detection.

\section{Generative Model to Detect DDoS Attacks: \\ A Use Case}\label{sec:methodology}

This section presents a use case that narrows the detection task, focusing on a binary task between network flows representing benign network communication versus DDoS attacks. 

We use the terminology \textit{flow} as the five-tuple (Source IP, Source Port, Destination IP, Destination Port, and Protocol) and the respective network packets in a given period. Our experiments use network flows from NetFlow-based datasets Bot-IoT, CICIDS-2018, and TON-IoT~\cite{Sarhan2022a}. These datasets comprise DDoS attacks that involve large-scale UDP, TCP, and HTTP requests.  
Our primary goal in this study is to develop a technique for identifying attacks in a multi-domain scenario represented by various network environments. The chosen datasets capture distinct contextual nuances for this goal while sharing a common feature set~\cite{Sarhan2022a}. Our focus is on identifying potential DDoS attacks across this multi-domain scenario. Furthermore, we exclusively trained the models using benign data representing an anomaly detection setting, which, from an operational perspective, would be simpler for deployment.

Thus, we propose a method called Anomaly-Flow that allows the classification of anomalous flows and the generation of synthetic data. It permits information exchange with other entities through external models trained with this synthetic data. 

\begin{figure*}[htb]
  \centering
  \includegraphics[width=0.7\linewidth]{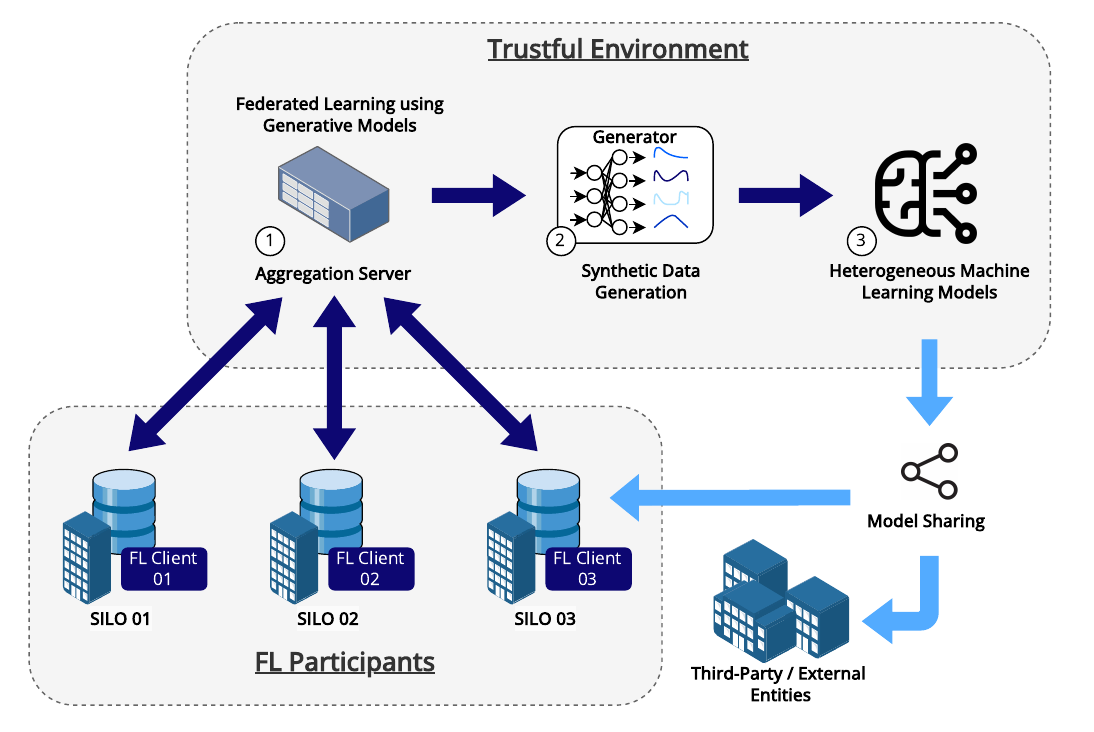}
  
  \caption{Use Case Training Diagram using Generative models in a FL schema \textcircled{\footnotesize{1}} FL using large datasets and the GANomaly model; \textcircled{\footnotesize{2}} Generation of synthetic data using the global model trained in the FL schema; \textcircled{\footnotesize{3}} Use the synthetic data generated by the generative model to train heterogeneous models that the FL participants and third-party entities can use.}
  \label{fig:proposed-solution}
\end{figure*}

The Anomaly-Flow training is presented in Fig. \ref{fig:proposed-solution}. In the first stage \ballnumberwhite{1}, an FL scheme is used to create a model capable of obtaining information from different data sources (participants’ silos) with privacy. 

Participants communicate with an aggregation server during the federated model’s training phase. The models are trained locally by FL clients for several epochs in each training round. In our study, each client trains for $50$ epochs during each round. After local training, each client sends only the model weights to the aggregation server. In turn, the server combines all the weights received from participants through an aggregation function, generating a global model. 

Finally, all local models are updated using the weights of the aggregated model, and the following rounds repeat the process until the specified number of rounds is reached. Furthermore, we use FedAvg as the aggregation algorithm. 

We use the GANomaly~\cite{ganomaly} algorithm as the federated model. This model uses a modified version of a GAN to identify image anomalies. However, this algorithm is also suitable for anomaly detection in different domains. 
A converged GANomaly model generates synthetic data resembling real samples by minimizing adversarial and contextual losses, which enforce feature consistency and contextual alignment with the training data. While theoretical alignment is achieved, further validation remains a potential direction for future work.

This work proposes changes for a network intrusion detection system (NIDS) based on network flows. Those changes are from the original GANomaly’s convolutional layers to dense layers, aligned with the tabular format of network flows as input.
Additionally, the \textit{generator} has decoder-dense layers with the structure of $256:512:1024$ and an encoder of the structure $1024:512:256$ — the \textit{discriminator} with layers of $1024:512:256$ and using a sigmoid function as output. A semi-supervised GAN network is appropriate for this work because it is based on anomaly detection and trained only using benign data. Companies deploying NIDS face challenges when attack data is unavailable, but the proposed model addresses this by training solely on benign data obtainable with contributions from network administrators and experts.

 In addition, for each participant in the trained model, dataset-specific rescaling information was calculated and kept local during training. Thresholds based on anomaly scores were used to classify examples: values below the threshold were labeled benign, while higher values indicated DDoS samples.

Next, we propose generating synthetic data with the collaboratively trained GAN in the method’s second stage \ballnumberwhite{2} -- referred to as \textit{Generator} on Fig~\ref{fig:proposed-solution}. This synthetic data aims to train heterogeneous models (diverse ML algorithms), allowing information sharing even to entities outside the FL scheme. 
Synthetic data enables the sharing of network context while preserving privacy, as the generated data, though related to actual data, represents non-existent flows. 
As the models are trained only with benign data, synthetic data consequently also represents a scenario of benign behavior. Furthermore, it should be noted that we consider that all use of synthetic data will be within a trustful perimeter and that it will not be externally accessible.

Finally, in the third \ballnumberwhite{3} and last stage of the proposed method, 
heterogeneous ML models can be trained with this synthetic data, simultaneously sharing the context of distinct participants. Besides, when using the synthetic data, we allow the training of models different from the one used in the FL setup. Therefore, external models become more flexible, enabling, for example, the use of lightweight models that are more suitable for resource-constrained devices. Moreover, we can share these models with external entities, maintaining the sensitive data within a trusting perimeter. For instance, a less complex algorithm like a tree-based model, which is simpler than a GAN, can be trained on this synthetic data and shared with third parties.

\subsection{Data Processing}\label{sec:data-procesing}

We preprocess the datasets for model training to remove inconsistencies and improve data organization. Initially, we split columns representing specific data protocol flags into multiple columns, each representing a particular value. Furthermore, the data pre-processing step removes attributes that bias training, such as IP address information and ports. Moreover, we remove examples containing outliers and null values. Following this, we divide benign samples from the datasets into three subsets: training (80 percent), testing (18 percent), and validation (2 percent). The validation subset determined the classification threshold using anomaly scores, which were then applied to evaluate the model during the test phase.

\subsection{Evaluation Metrics}

\begin{figure}[htb]
    \centering
    \includegraphics[width=1\linewidth]{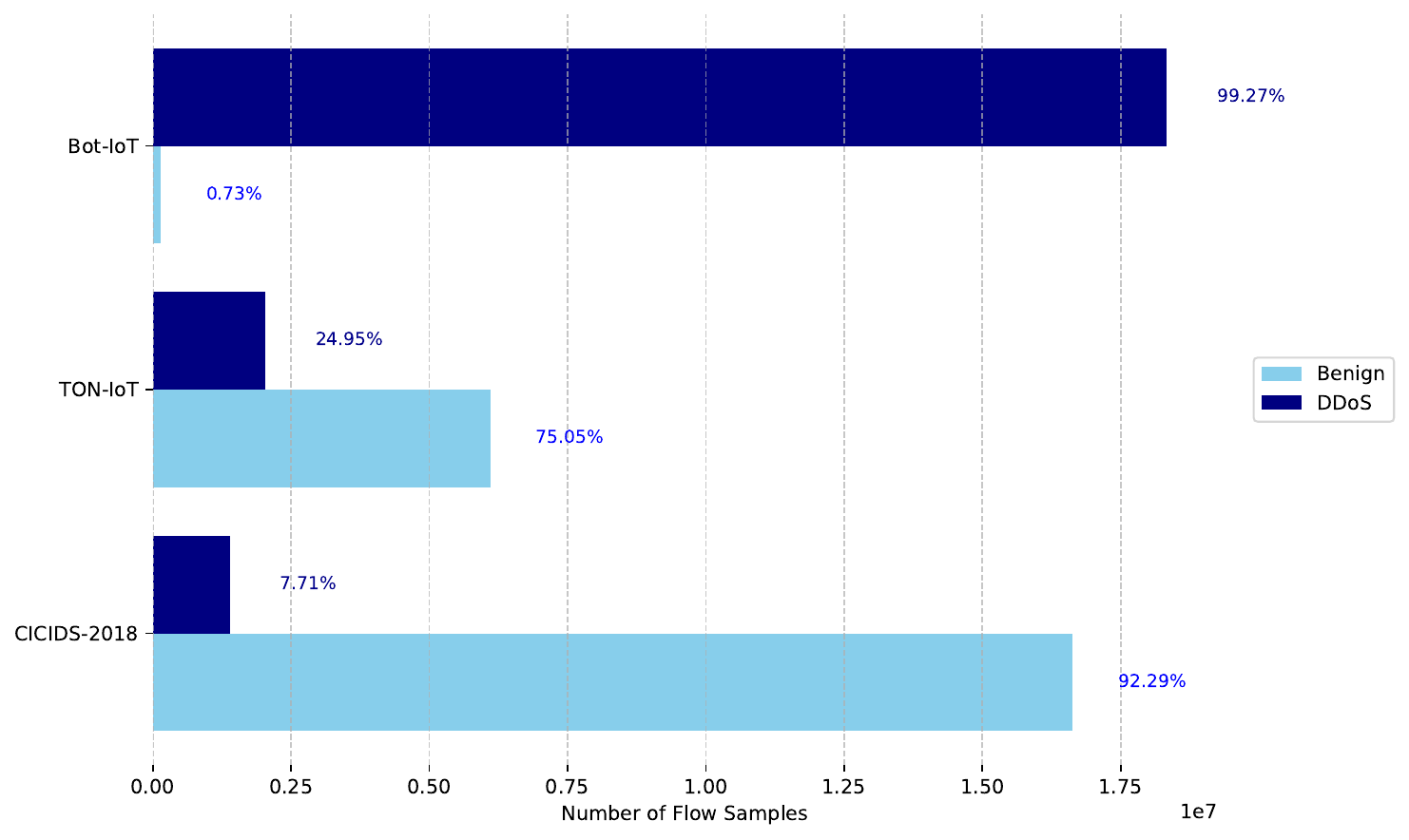}
    \caption{Class distribution and imbalance across analyzed datasets for DDoS detection, highlighting the skewness in class proportions and sample quantities. The benign class dominates in CICIDS-2018 and TON-IoT, whereas DDoS traffic overwhelmingly prevails in Bot-IoT, illustrating significant disparities in data distribution across datasets.}
    \label{fig:balancing}
\end{figure}

Our experiments evaluate the models’ performance using two learning metrics: Area Under the Curve of the Receiver Operation Curve (ROC-AUC) and F1-score. 
Such metrics are due to the characteristic of the problem that presents an imbalanced number of samples in the classes (benign and DDoS, as reported in Fig.~\ref{fig:balancing}) and the strategy using an anomaly score from GANomaly for each sample (continuous real value between zero and one), which allows the threshold evaluation to compose the ROC curve. 

\subsection{Cross-Evaluation} 
\label{sub-section-cross}
\begin{figure}[htb] 
    \centering
  \subfloat[Training Model / Testing Model Locally\label{fig:cross-evaluation-a}]{%
       \includegraphics[width=0.65\linewidth]{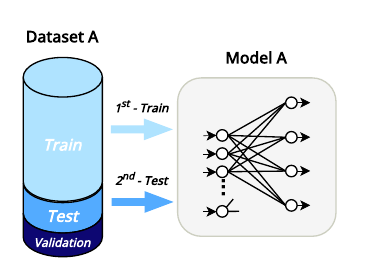}}
    \hfill
  \subfloat[Cross-Evaluation Test\label{fig:cross-evaluation-b}]{%
        \includegraphics[width=0.65\linewidth]{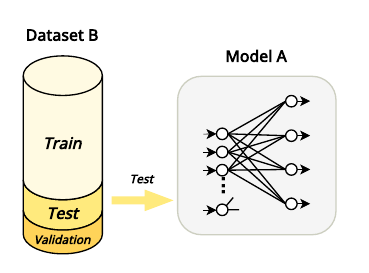}}
  \caption{(a) The diagram presents the structure of the data split for the training and test. Initially, model A is trained on the train split of dataset A and then evaluated with the test set from the same dataset used in training, referred to as local evaluation. Next, (b) presents the cross-evaluation procedure, in which model A, previously trained on dataset A, is evaluated with data from a different dataset, in this example, dataset B.}
  \label{fig:cross-evaluation} 
\end{figure}

We employ a cross-evaluation methodology to comprehensively assess the model’s performance and multi-domain capabilities, as shown in Fig. \ref{fig:cross-evaluation}. Initially, the model undergoes local assessment, training, and testing on the same dataset (Fig. \ref{fig:cross-evaluation-a}). After that, a cross-evaluation is conducted, entailing training the model on one dataset and testing it on another (Fig. \ref{fig:cross-evaluation-b}).

This cross-evaluation is paramount in discerning the model’s ability to identify previously unseen attacks and gauging the variations in its performance across different datasets representing a multi-domain setting. The initial local evaluation is a comparative baseline for the subsequent federated evaluation. Subsequently, a single FL model is evaluated in multiple contexts to assert the generalization of the proposed model.

\section{Results and Discussion}\label{sec:results}

This section presents the results obtained during the execution of the experiments to evaluate Anomaly-Flow. We present the results and discussion in three steps: developing a GAN model in an FL setup to learn in a cross-domain setting (generalization capability), generating synthetic flows using GAN, and evaluating the sharing of GAN’s synthetic data for external model training. 

Firstly, we present the results related to \ballnumberwhite{1} of our proposed solution regarding the generalization capability. 

Next, we present the results of \ballnumberwhite{2} and \ballnumberwhite{3} about synthetic data generation and training different models using the generated examples. Furthermore, we present a discussion along with the results.

\subsection{Multi-domain DDoS Detection} 
\label{sec:gen-capacity} 

The performance of the models and the generalization capability in a multi-domain scenario used the learning metrics ROC-AUC and F1-score. 

We present the evaluation results for each test dataset in Table \ref{tab:performance-eval}. A GANomaly-based model was trained on each dataset listed in the Model column. Then, the trained models were assessed using their data (test split) and against the other datasets with the same features (i.e., cross-evaluation). Lastly, in an FL setup and after ten training rounds, the global GANomaly-based model was evaluated against the three datasets, and its metrics are also present in Table \ref{tab:performance-eval} representing \ballnumberwhite{1}.

\begin{table*}[htbp]
\caption{Performance Measurement for Local and Cross-Evaluation of GANomaly Models vs. GANomaly on a FL setup after 10 rounds, considering no data shared among silos. The first three models in the first column represent “trained on” (CICIDS-2018, Bot-IoT, TON-IoT), and the same dataset name in the Evaluation Dataset column means the evaluation in the same dataset (no cross-evaluation), and the other two datasets (cross-evaluation representing a multi-domain setting). Lastly, the FL result after ten rounds.}
\label{tab:performance-eval}
\centering
\resizebox{0.70\textwidth}{!}{
\begin{tabular}{llcc}
\toprule
\multicolumn{1}{l}{\textbf{Model}}                 & \multicolumn{1}{l}{\textbf{Evaluation dataset}} & \multicolumn{1}{l}{\textbf{ROC-AUC}}  & \multicolumn{1}{l}{\textbf{F1-score}} \\ \midrule

\multicolumn{1}{l}{\multirow{3}{*}{GANomaly trained on CICIDS-2018}} & \multicolumn{1}{l}{CICIDS-2018}    & \multicolumn{1}{r}{0.924}               & \multicolumn{1}{r}{0.615}         \\ %\cline{2-6} 
\multicolumn{1}{l}{}                                 & \multicolumn{1}{l}{Bot-IoT}            & \multicolumn{1}{r}{0.892}               & \multicolumn{1}{r}{0.989}         \\ %\cline{2-6} 
\multicolumn{1}{l}{}                                 & \multicolumn{1}{l}{TON-IoT}            & \multicolumn{1}{r}{0.885}               & \multicolumn{1}{r}{0.751}         \\ \midrule
\multicolumn{1}{l}{\multirow{3}{*}{GANomaly trained on Bot-IoT}}         & \multicolumn{1}{l}{Bot-IoT}            & \multicolumn{1}{r}{0.808}               & \multicolumn{1}{r}{0.998}         \\ % \cline{2-6} 
\multicolumn{1}{l}{}                                 & \multicolumn{1}{l}{CICIDS-2018}    & \multicolumn{1}{r}{0.419}               & \multicolumn{1}{r}{0.062}         \\ % \cline{2-6} 
\multicolumn{1}{l}{}                                 & \multicolumn{1}{l}{TON-IoT}            & \multicolumn{1}{r}{0.474}               & \multicolumn{1}{r}{0.743}         \\ \midrule
\multicolumn{1}{l}{\multirow{3}{*}{GANomaly trained on TON-IoT}}         & \multicolumn{1}{l}{TON-IoT}    & \multicolumn{1}{r}{0.891}               & \multicolumn{1}{r}{ 0.867}         \\ % \cline{2-6} 
\multicolumn{1}{l}{}                                 & \multicolumn{1}{l}{Bot-IoT}            & \multicolumn{1}{r}{0.691}               & \multicolumn{1}{r}{0.998}         \\ % \cline{2-6} 
\multicolumn{1}{l}{}                                 & \multicolumn{1}{l}{CICIDS-2018}            & \multicolumn{1}{r}{0.764}               & \multicolumn{1}{r}{ 0.147 }         \\ \midrule %\hline \hline

\multicolumn{1}{l}{\multirow{3}{*}{GANomaly on a FL setup}}         & \multicolumn{1}{l}{CICIDS-2018}    & \multicolumn{1}{r}{0.926}               & \multicolumn{1}{r}{0.493}         \\ % \cline{2-6} 
\multicolumn{1}{l}{}                                 & \multicolumn{1}{l}{Bot-IoT}            & \multicolumn{1}{r}{0.994}               & \multicolumn{1}{r}{0.968}         \\ % \cline{2-6} 
\multicolumn{1}{l}{}                                 & \multicolumn{1}{l}{TON-IoT}            & \multicolumn{1}{r}{0.866}               & \multicolumn{1}{r}{0.781}         \\

\bottomrule
\end{tabular}}
\end{table*}

Table~\ref{tab:baselines-to-anomalyflow} compares Anomaly-Flow with baselines for multi-domain DDoS detection. We used the architecture of \cite{cose} for the AE comparison, excluding the proposed stacking architecture and dual-threshold mechanism. We evaluated the AE solely for benign versus DDoS anomaly detection. However, this simple AE did not achieve satisfactory performance, as indicated by the reported F1-score. 
\cite{globecom-ddos} introduces a basic logistic regression model within a federated learning (FL) framework for multi-domain DDoS detection. Nevertheless, its performance in multi-domain evaluations is lower than that of Anomaly-Flow, based on the same reference datasets. FLAD~\cite{FLAD2024} aims to improve convergence in FL-based DDoS detection, but does not explicitly evaluate multi-domain detection. In this study, we extend FLAD’s evaluation to the same multi-domain detection scenario as proposed by Anomaly-Flow, but FLAD still underperforms relative to Anomaly-Flow. We did not explore FLAD’s client-specific optimization, 
as this is not part of the Anomaly-Flow methodology.

{
\begin{table*}[htbp]
\caption{Comparison of Anomaly-Flow and baselines in a federated learning setting, focusing on multi-domain DDoS detection by average F1-score. The baselines are evaluated in a cross-evaluation setting, with CICIDS-2018, Bot-IoT, and TON-IoT representing the multi-domain scenario.}
\label{tab:baselines-to-anomalyflow}
\centering

\resizebox{\textwidth}{!}{%
\begin{tabular}{llr}

\toprule
\textbf{Reference}  & \textbf{Scope}  & \textbf{Average F1-score} \\ 
\midrule
Autoencoder & Multi-domain detection with FL (excluding \cite{cose} changes, adapted for DDoS) & $0.451 \pm 0.329$ \\

Logistic Regression~\cite{globecom-ddos} & DDoS multi-domain detection with FL & $0.496 \pm 0.097$ \\

FLAD~\cite{FLAD2024} & Improving FL Convergence for DDoS & $0.694 \pm 0.336$            \\

Anomaly-Flow & DDoS multi-domain detection and Pre-trained Models using Synthetic Data & $0.747 \pm 0.195$            \\ \bottomrule
\end{tabular}%
}
\end{table*}
}

\begin{figure*}[htb]
    \centering
    \includegraphics[width=0.8\linewidth]{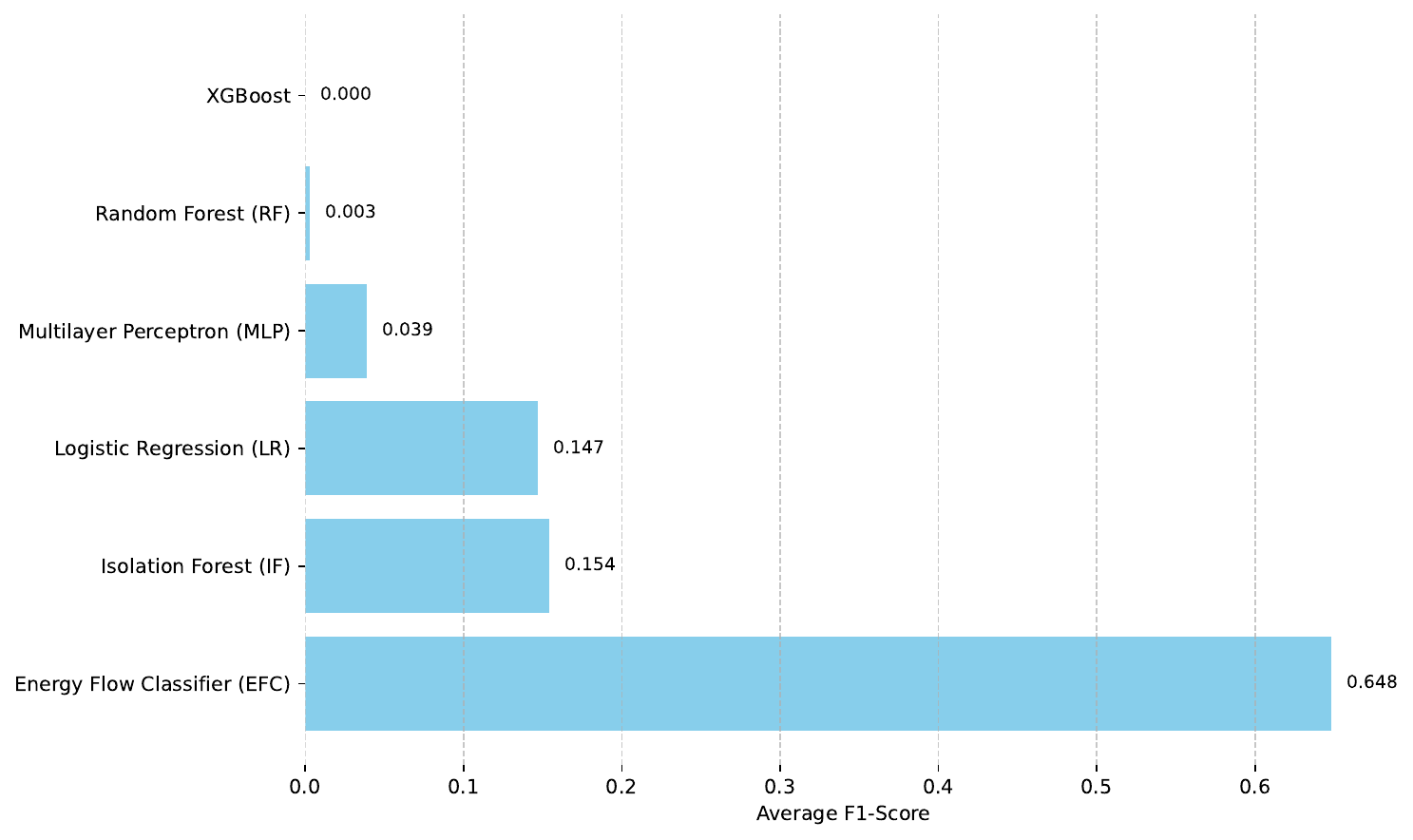}
    \caption{Average F1-score for baseline algorithms, showing their generalization performance when trained on one dataset and tested on the other two in the context of DDoS detection. The results reveal significant variability in performance across models, with most algorithms achieving low scores, indicating poor cross-dataset generalization. Notably, the Energy Flow Classifier (EFC) substantially outperforms others, suggesting its robustness in diverse network scenarios. These findings emphasize the challenges of deploying machine learning models in heterogeneous environments and underscore the importance of designing algorithms capable of handling such variability.}
    \label{fig:baselines}
\end{figure*}

Furthermore, to establish a baseline for comparison with other models, we performed experiments using various ML algorithms under the same data processing and separation conditions described in Section~\ref{sec:data-procesing}. We cross-evaluated these various ML algorithms as presented in Fig. \ref{fig:cross-evaluation}.  
We only considered the average F1-score as a reference for these baselines. In Fig.~\ref{fig:baselines}, we have reported the average F1-score metric values obtained for each baseline model.

The results in different contexts using the cross-evaluation, compared with similar studies conducted previously, demonstrate the robustness of the GANomaly models. Comparatively, using the average F1-score, it is possible to compare the performance of this work using GANomaly, which was $0.615$ (averaging the six cross-evaluation metrics from CICIDS-2018, Bot-IoT and TON-IoT), representing a considerable result in the ability to classify DDoS attacks. This result is $5.1\%$ lower than the best baseline obtained with the EFC algorithm (Fig.~\ref{fig:baselines}). 

However, the EFC algorithm needs to be trained with all the data each time, so if it is necessary to train the models with new data, it is required to recover the previous training data, thus making incremental training of the models cumbersome. 

Considering the FL task for GANomaly -- GANomaly on an FL setup from Table~\ref{tab:performance-eval} -- the metrics’ average demonstrates that after $10$ rounds, it can detect DDoS better than the previous approaches using GANomaly on each dataset without FL. 
The average F1-score obtained is $0.747$, based on the evaluation of each of the participants (CICIDS-2018, Bot-IoT, TON-IoT). Although the model presented reaches a sub-optimal point, it can reasonably identify attacks. Considering a real scenario, this result represents a mitigation of DDoS attacks in a multi-domain setting, assuming the application of the same model to different contexts. 

Using models trained on a specific context data yields good results within that context. However, performance significantly degrades when evaluated in other contexts, as reported in Fig.~\ref{fig:baselines} for most algorithms. For example, in Table~\ref{tab:performance-eval}, consider evaluating the GANomaly trained with the Bot-IoT dataset: within its context (its test set), it achieved an F1-score of $0.998$. However, when assessed in a different context, such as CICIDS-2018, its performance dropped to $0.062$. This demonstrates the model’s limited effectiveness from a multi-domain perspective.

 Considering the ROC-AUC values in Table \ref{tab:performance-eval},  the GANomaly trained on CICIDS-2018 and TON-IoT datasets obtained a better cross-evaluation result than the Bot-IoT case. 
 A factor in this difference is the amount of benign and DDoS data in each dataset; this measure is presented in Fig.~\ref{fig:balancing}. In this way, the Bot-IoT dataset is highly imbalanced, containing a low percentage of samples referring to benign flows. This factor may have influenced the training of the models and, consequently, its cross-evaluation performance.

\subsection{External Models Evaluation} 
\label{sec:external-models}

The external models’ evaluation analysis uses an additional dataset (UNSW-NB15) representing an external entity. We trained different heterogeneous models using synthetic data \ballnumberwhite{2} and shared those models \ballnumberwhite{3} with the external entity without taking data outside the trustful environment (Fig.~\ref{fig:proposed-solution}). Thus, sharing a model instead of synthetic data adds another layer of privacy. The shared model \ballnumberwhite{3} was trained with benign synthetic network data, which enables the external entity to start from this previous knowledge. Then, the external entity can fine-tune this shared model with its data to adapt to its domain. Next, our analysis is two-fold: how this shared model performs on the external entity and whether this external entity identifies anomalous DDoS samples never seen before.

We considered as external models the algorithms capable of incremental learning: Random Forest, Isolation Forest, XGBoost, and MLP -- EFC and LR are unfeasible for incremental learning. The models were trained using 100,000 synthetic samples. Next, the models were trained in this new domain incrementally based on UNSW-NB15 data. Then, we evaluated the model’s capability to detect attacks on this new domain. On the one hand, the algorithms Isolation Forest (F1-score: 0.018) and XGBoost (F1-score: 0.076) performed insufficiently. On the other hand, the algorithms Random Forest (F1-score: 0.965) and MLP (F1-score: 0.940) obtained reasonable results in the new domain data after fine-tuning.

Subsequently, we used TON-IoT, CICIDS-2018, and Bot-IoT test data to evaluate this model’s performance when facing DDoS samples that were never seen before in its domain (UNSW-NB15). The results obtained in this analysis were the following: the MLP algorithm is the most promising, detecting DDoS samples from TON-IoT and Bot-IoT, achieving an F1-score of 0.750 and 0.989, respectively. However, CICIDS-2018 achieved an F1-score of 0.024, a possible reason is that the synthetic data could be statistically distant from the CICIDS-2018 domain. Overall, the external model could identify DDoS attacks from other domains.

The results of sharing synthetic data as a pre-trained model were behind expectations. However, when sharing synthetic data, the model began to identify previously unseen DDoS samples from other domains. This result highlights challenges that still need to be overcome, such as the semantic quality of the data generated by GAN networks. It is also noteworthy that sharing attacks and benign data could improve the models’ performance. Furthermore, by refining the presented proposal, other methods can be applied to enhance the performance of models trained with synthetic data.

\section{Challenges}
\label{sec:challenges}

Using machine learning for NIDS and techniques to enhance generalization capabilities still pose challenges. The following are challenges identified regarding the approach to those topics.

\textbf{Generalization across domains:} Different networks exhibit varying types of traffic during use. For instance, IoT networks have significant differences compared to corporate networks. Furthermore, DDoS attacks also have specific traits depending on the tool used to generate the attacks. Considering the datasets analyzed in this work, different DoS attack tools exist in each dataset on which the models were trained. Each dataset involved a range of DDoS tools, highlighting the diversity of attack generation methods. 

\textbf{Data sharing privacy}: Sharing information between multiple networks can be a way to identify a greater variety of attacks. However, information sharing must consider several factors, such as sharing sensitive information or corporate secrets. The use of FL contributes to sharing information relating only to machine learning models. However, there are still concerns regarding attacks on the FL system. We propose sharing synthetic data via a pre-trained model to enable collaborative NIDS while preserving privacy, though it demands extra effort.

\textbf{Data Quality, Availability, and Updates}: It becomes challenging for the developer to identify whether the data used for training has all the valid information and also for the synthetically generated data. Given this uncertainty, the models can be trained incorrectly and represent a behavior different from the applications’ regular use.

For instance, during our evaluation of Anomaly-Flow’s synthetic generated data, initial findings about the feature \textit{protocol}, which would be expected to range between $0$ and $255$, for some synthetic flows this feature assumed values above $4000$, which reinforces this challenge. Furthermore, data drift in network data is intense, as factors such as seasonal operations, new applications, and obsolete detection systems require constant retraining with new information.

In addition, regarding quality, datasets must have information from attacks that have a substantial impact. Therefore, providing the ability to update data and models is an open challenge for NIDS research.

\textbf{Deployment and use of ML NIDS in real-world}: Many works present solutions based on ML, but adopting the tools in a corporate environment is still a long way off. Application in practice requires that models be capable of identifying attacks without generating many false alarms. Erroneous attributions result in time-consuming analyses and generate unnecessary costs for operators. It still becomes an even more significant challenge to explain the identification quickly and generate information to facilitate future studies.

\section{Opportunities}
\label{sec:opportunities}

Identifying DDoS attacks in a multi-domain setting presents opportunities for improvement concerning the challenges presented. The identification of attacks can be carried out through the identification of anomalies but without the use of fixed thresholds. The use of adaptive thresholds or even the composition with machine learning systems can contribute to improving the performance of models, taking into account differences in the data in each context. 

From a research perspective, cross-evaluation between different datasets should be an essential component of future research to validate whether reported results generalize across scenarios and to document limitations explicitly. This rigorous evaluation approach would help bridge the gap between theoretical advances and practical implementations. Additionally, it would provide more precise insight into which detection methods are genuinely effective across different network environments.

Researchers can explore techniques to enhance the semantic quality of synthetic data by evaluating it with domain experts and implementing constraints that prevent inconsistent conditions (e.g., negative time measurements) while preserving the data's utility for detection tasks. 
Creating a method to identify new DDoS attack tools and techniques is necessary regarding data quality, availability, and updates. When placing new participants, applying a method capable of grouping attack examples with benign samples is essential, thus creating datasets that represent potentially harmful samples. Finally, developing a tool capable of simulating environments as closely as possible to reality is critical. Including experts' feedback to ensure that using the NIDS system becomes practical is also an opportunity.

\section{Conclusions}
\label{sec:conclusion}

This work demonstrated how GAN networks configured in an FL scheme can help identify DDoS in a multi-domain setting. The proposed GAN system can generate synthetic data based on FL participants. We presented that heterogeneous models trained using this synthetic data can identify previously unseen DDoS attacks. 

However, the solution’s performance requires more effort to be implemented and used in practice by network operators. Thus, we report several challenges encountered during our tests. The first challenge is that generalization across domains remains an open issue despite the advancements achieved through our approach.

Furthermore, the performance of heterogeneous models is highly dependent on the data generated synthetically, so the distribution of the generated data must be as close to reality as possible. Moreover, the operational aspects of the models must be considered, such as the implementation of the system, stages of updating training data, and the models’ ability to support decision-making. 

Finally, we list research opportunities for overcoming these challenges, serving as future directions for research in the area.

The code to reproduce the experiments is available in the repository \url{https://github.com/c2dc/anomaly-flow}.

%TC:ignore
\section*{Acknowledgments}

Work supported in part by ITA’s Programa de Pós-gradua\c{c}ão em Aplica\c{c}ões Operacionais (ITA/PPGAO).  The authors are partially supported by the grant \#2020/09850-0 and \#2022/00741-0. São Paulo Research Foundation (FAPESP) and CAPES.
%TC:endignore

\bibliographystyle{unsrt}
\bibliography{bibfile}

\begin{thebibliography}{10}

\bibitem{generalization-survey}
Jindong Wang, Cuiling Lan, Chang Liu, Yidong Ouyang, Tao Qin, Wang Lu, Yiqiang
  Chen, Wenjun Zeng, and Philip~S. Yu.
\newblock Generalizing to unseen domains: A survey on domain generalization.
\newblock {\em IEEE Transactions on Knowledge and Data Engineering},
  35(8):8052--8072, 2023.

\bibitem{AGRAWAL2022346}
Shaashwat Agrawal, Sagnik Sarkar, Ons Aouedi, Gokul Yenduri, Kandaraj Piamrat,
  Mamoun Alazab, Sweta Bhattacharya, Praveen Kumar~Reddy Maddikunta, and
  Thippa~Reddy Gadekallu.
\newblock Federated learning for intrusion detection system: Concepts,
  challenges and future directions.
\newblock {\em Computer Communications}, 195:346--361, 2022.

\bibitem{flddos}
Jiachao Zhang, Peiran Yu, Le~Qi, Song Liu, Haiyu Zhang, and Jianzhong Zhang.
\newblock {FLDDoS: DDoS Attack Detection Model based on Federated Learning}.
\newblock In {\em 2021 IEEE 20th International Conference on Trust, Security
  and Privacy in Computing and Communications (TrustCom)}, pages 635--642,
  2021.

\bibitem{fleam}
Jianhua Li, Lingjuan Lyu, Ximeng Liu, Xuyun Zhang, and Xixiang Lyu.
\newblock {FLEAM: A Federated Learning Empowered Architecture to Mitigate DDoS
  in Industrial IoT}.
\newblock {\em IEEE Transactions on Industrial Informatics}, 18(6):4059--4068,
  2022.

\bibitem{FLAD2024}
Roberto Doriguzzi-Corin and Domenico Siracusa.
\newblock {FLAD: Adaptive Federated Learning for DDoS attack detection}.
\newblock {\em Computers \& Security}, 137:103597, 2024.

\bibitem{xenids}
Giovanni Apruzzese, Luca Pajola, and Mauro Conti.
\newblock The cross-evaluation of machine learning-based network intrusion
  detection systems.
\newblock {\em IEEE Transactions on Network and Service Management},
  19(4):5152--5169, 2022.

\bibitem{CATILLO2021102341}
Marta Catillo, Antonio Pecchia, Massimiliano Rak, and Umberto Villano.
\newblock {Demystifying the role of public intrusion datasets: A replication
  study of DoS network traffic data}.
\newblock {\em Computers \& Security}, 108:102341, 2021.

\bibitem{Verkerken2021}
Miel Verkerken, Laurens D'hooge, Tim Wauters, Bruno Volckaert, and Filip
  De~Turck.
\newblock Towards model generalization for intrusion detection: Unsupervised
  machine learning techniques.
\newblock {\em Journal of Network and Systems Management}, 30(1):12, Oct 2021.

\bibitem{cose}
Gustavo {de Carvalho Bertoli}, Lourenço {Alves Pereira Junior}, Osamu Saotome,
  and Aldri~Luiz {dos Santos}.
\newblock Generalizing intrusion detection for heterogeneous networks: A
  stacked-unsupervised federated learning approach.
\newblock {\em Computers \& Security}, 127:103106, 2023.

\bibitem{netshare}
Yucheng Yin, Zinan Lin, Minhao Jin, Giulia Fanti, and Vyas Sekar.
\newblock Practical gan-based synthetic ip header trace generation using
  netshare.
\newblock In {\em Proceedings of the ACM SIGCOMM 2022 Conference}, SIGCOMM '22,
  page 458–472, New York, NY, USA, 2022. Association for Computing Machinery.

\bibitem{PST2022}
Euclides Carlos~Pinto Neto, Sajjad Dadkhah, and Ali~A. Ghorbani.
\newblock {Collaborative DDoS Detection in Distributed Multi-Tenant IoT using
  Federated Learning}.
\newblock In {\em 2022 19th Annual International Conference on Privacy,
  Security \& Trust (PST)}, pages 1--10, 2022.

\bibitem{efc}
Camila F.~T. Pontes, Manuela M.~C. de~Souza, João J.~C. Gondim, Matt Bishop,
  and Marcelo~Antonio Marotta.
\newblock A new method for flow-based network intrusion detection using the
  inverse potts model.
\newblock {\em IEEE Transactions on Network and Service Management},
  18(2):1125--1136, 2021.

\bibitem{Sarhan2022a}
Mohanad Sarhan, Siamak Layeghy, and Marius Portmann.
\newblock Towards a standard feature set for network intrusion detection system
  datasets.
\newblock {\em Mobile Networks and Applications}, 27(1):357--370, Feb 2022.

\bibitem{ganomaly}
Samet Akcay, Amir Atapour-Abarghouei, and Toby~P Breckon.
\newblock Ganomaly: Semi-supervised anomaly detection via adversarial training.
\newblock In {\em Asian Conference on Computer Vision}, pages 622--637.
  Springer, 2018.

\bibitem{globecom-ddos}
Leonardo~H de~Melo, Gustavo de~C~Bertoli, Lourenco~A Pereira, Osamu Saotome,
  Marcelo~F Domingues, and Aldri~Luiz dos Santos.
\newblock {Generalizing Flow Classification for Distributed Denial-of-Service
  over Different Networks}.
\newblock In {\em {GLOBECOM 2022 - 2022 IEEE Global Communications
  Conference}}, pages 879--884, 2022.

\end{thebibliography}

\end{document}